\documentclass[10pt]{article}
\usepackage{graphicx}
\usepackage{amsmath}
\usepackage{amssymb}
\usepackage{caption2}
\begin{document}
\begin{center}
{\large {\bf \sc{   Analysis of the   triply-charmed pentaquark states  with QCD sum rules
  }}} \\[2mm]
Zhi-Gang  Wang \footnote{E-mail: zgwang@aliyun.com.  }   \\
 Department of Physics, North China Electric Power University, Baoding 071003, P. R. China
\end{center}

\begin{abstract}
In this article, we construct the scalar-diquark-scalar-diquark-antiquark type current to study the  ground state triply-charmed  pentaquark states with the QCD sum rules. We  separate the contributions of the negative-parity and positive-parity triply-charmed  pentaquark states explicitly, and take the energy scale formula $\mu=\sqrt{M^2_{P}-(3{\mathbb{M}}_c)^2}$ to determine the optimal energy scales of the QCD spectral densities. The predicted pentaquark masses can be confronted to the experimental data in the future.
\end{abstract}



\section{Introduction}
   The  diquarks $\varepsilon^{ijk} q^{T}_j C\Gamma q^{\prime}_k$
  have  five  structures  in Dirac spinor space, where  $C\Gamma=C\gamma_5$, $C$, $C\gamma_\mu \gamma_5$,  $C\gamma_\mu $ and $C\sigma_{\mu\nu}$ for the scalar, pseudoscalar, vector, axialvector  and  tensor diquarks, respectively, the $i$, $j$, $k$  are color indexes.  The
attractive interaction of one-gluon exchange  favors  formation of
the diquarks in  color antitriplet $\overline{3}_{ c}$, flavor
antitriplet $\overline{3}_{ f}$ and spin singlet $1_s$ or flavor
sextet  $6_{ f}$ and spin triplet $3_s$ \cite{One-gluon}.
   The calculations based on the QCD sum rules  indicate that  the favored configurations are the $C\gamma_5$ and $C\gamma_\mu$ diquark states \cite{WangDiquark,WangLDiquark,TangHH}, the light-light $\varepsilon^{ijk} q^{T}_j C\gamma_5 q^{\prime}_k$ diquark states have much smaller masses  than the corresponding $\varepsilon^{ijk} q^{T}_j C\gamma_\mu q^{\prime}_k$ diquark states \cite{WangLDiquark}, while the heavy-light or heavy-heavy $\varepsilon^{ijk} q^{T}_j C\gamma_5 q^{\prime}_k$ and $\varepsilon^{ijk} q^{T}_j C\gamma_\mu q^{\prime}_k$ diquark states have almost  degenerate masses  \cite{WangDiquark,TangHH}. All in all, the lowest
   states are the scalar diquark states, although the energy gaps between the scalar and axialvector diquark states are rather small in some cases.
   We can construct the lowest tetraquark states, pentaquark states and hexaquark states with  the  $C\gamma_5$ and $C\gamma_\mu$ diquark states or antidiquark states.
   Experimentally, the  $Z^\pm_c(3900)$ and $Z^{\pm}_c(4020/4025)$ observed by the BESIII collaboration  \cite{BES3900,BES1308,BES1309}, the $Z^{\pm}_b(10610)$, $Z^{\pm}_b(10650)$, $Z_c(4200)^\pm$ observed by  the Belle collaboration \cite{Belle1110,Zc4200exp,YuanCZ4200}, the $Z_c^\pm(4430)$ observed by the  Belle collaboration \cite{Belle-2007} and confirmed by the LHCb collaboration \cite{LHCb-1404}, the $P_c^+(4380)$ and $P_c^+(4450)$ observed by  the  LHCb collaboration   \cite{LHCb-4380}, provide excellent candidates for the hidden-charm or hidden-bottom tetraquark states and pentaquark states.

The QCD sum rules is a powerful nonperturbative  tool in studying the ground state hadrons, and has  given many successful descriptions of
  the hadronic parameters on the phenomenological side \cite{SVZ79,Reinders85}.  For example, the $Z_c^\pm(3900)$
can be tentatively assigned to be the ground state  $C\gamma_5\otimes \gamma_\mu C-C\gamma_\mu\otimes\gamma_5 C$ type tetraquark state \cite{WangHuang-3900,Wang-4430,Nielsen3900} or the $i\gamma_5\otimes \gamma_\mu +\gamma_\mu\otimes i\gamma_5 $ type molecular state \cite{WangMolecule-3900, ZhangJR-3900}.
In Ref.\cite{WangHuang-3900}, we tentatively assign the $X(3872)$ and $Z_c^\pm(3900)$ to be the axialvector tetraquark states and study their masses with the QCD sum rules in a systematic way, and explore the energy scale dependence of the hidden-charm tetraquark states  in details for the first time.
In Ref.\cite{Wang-4660-2014}, we study the diquark-antidiquark type hidden-charm vector tetraquark states in details and suggest a  formula,
\begin{eqnarray}
\mu&=&\sqrt{M^2_{X/Y/Z}-(2{\mathbb{M}}_c)^2} \, ,
 \end{eqnarray}
 with the effective $c$-quark mass ${\mathbb{M}}_c$ to determine the optimal energy scales of the  QCD spectral densities in the QCD sum rules. The formula also works well for the diquark-diquark-antiquark type hidden-charm pentaquark states \cite{WangPc}, and  be extended to study the diquark-diquark-diquark type doubly-charmed hexaquark state to enhance the pole contribution \cite{Wang-HexaQuark}.

In 2017, the LHCb collaboration observed the doubly-charmed baryon state  $\Xi_{cc}^{++}$ in the $\Lambda_c^+ K^- \pi^+\pi^+$ mass spectrum  \cite{LHCb-Xicc}.
The doubly heavy tetraquark state $QQ\bar{q}\bar{q}^\prime$ is very similar to the doubly heavy baryon state $QQq$,
 where  we have a light antidiquark  $\bar{q}\bar{q}^\prime$ instead of a light quark $q$ in color triplet. The energy scale formula $\mu=\sqrt{M^2_{X/Y/Z}-(2{\mathbb{M}}_Q)^2} $ also works well for the doubly heavy tetraquark states \cite{Wang-cc-tetraquark}. Recently, the triply heavy  tetraquark states were studied in detailed with the QCD sum rules \cite{Zhu-ccc-tetraquark}. So it is interesting to study the triply heavy pentaquark states with the QCD sum rules. At the first step, we study the triply-charmed pentaquark states and explore the energy scale dependence of the QCD spectral densities. The diquark-diquark-antiquark type triply-charmed pentaquark states differ from the baryon-meson type triply-charmed molecular states remarkably, which have been studied with  the
 potential models based on the heavy quark symmetry \cite{Guo-Liu}.

 In this article, we choose the  $ u^{T} C\gamma_{5} c- d^{T} C\gamma_{5} c-\bar{c}$ type configuration  to study the lowest $cc\bar{c}ud$ pentaquark state with $J^P={\frac{1}{2}}^-$ by calculating the operator product expansion up to dimension 10 and extend the energy scale formula $\mu=\sqrt{M^2_{X/Y/Z}-(2{\mathbb{M}}_c)^2}$ to the new form $\mu=\sqrt{M^2_{P}-(3{\mathbb{M}}_c)^2}$ to determine the ideal energy scale of the QCD spectral density of the triply-charmed pentaquark state, as a byproduct, we also study the $J^P={\frac{1}{2}}^+$ $cc\bar{c}ud$ pentaquark state with QCD sum rules. One may expect to study
  the lowest $cc\bar{c}ud$ pentaquark state with the configuration $ c^{T} C\gamma_{\mu} c- u^{T} C\gamma^{\mu} d-\bar{c}$. Naively, we expect that the larger masses of the $C\gamma_{\mu}$ diquark states lead to larger tetraquark or pentaquark masses compared to the $C\gamma_5$ diquark states.

The article is arranged as follows:  we derive the QCD sum rules for the masses and pole residues of  the
 triply-charmed pentaquark states in Sect.2;  in Sect.3, we present the numerical results and discussions; and Sect.4 is reserved for our
conclusion.

\section{QCD sum rules for  the $ {\frac{1}{2}}^{\pm}$ pentaquark states}

In the following, we write down  the two-point correlation function $\Pi(p^2)$  in the QCD sum rules,
\begin{eqnarray}
\Pi(p^2)&=&i\int d^4x e^{ip \cdot x} \langle0|T\left\{J(x)\bar{J}(0)\right\}|0\rangle \, ,
\end{eqnarray}
where
 \begin{eqnarray}
 J(x)&=&\varepsilon^{ila} \varepsilon^{ijk}\varepsilon^{lmn}  u^T_j(x) C\gamma_5 c_k(x)\,d^T_m(x) C\gamma_5 c_n(x)\,   C\bar{c}^{T}_{a}(x) \, ,
\end{eqnarray}
 the $i$, $j$, $k$, $l$, $m$, $n$ and $a$ are color indexes, the $C$ is the charge conjugation matrix.
In this article, we   choose   the scalar-diquark-scalar-diquark-antiquark type  current $J(x)$ with $J^P={\frac{1}{2}}^{-}$
    to study the lowest triply-charmed  pentaquark states with $J^P={\frac{1}{2}}^{\pm}$ in a consistent way.

The current $J(0)$ has negative parity, and    couples potentially to the $J^P={\frac{1}{2}}^-$   triply-charmed  pentaquark  state $P^{-}$,
\begin{eqnarray}
\langle 0| J (0)|P^{-}(p)\rangle &=&\lambda_{-} U^{-}(p,s) \, ,
\end{eqnarray}
the $\lambda_{-}$ is the pole residue, the spinor $U^{-}(p,s)$   satisfies the Dirac equation $(\not\!\!p-M_{-})U^{-}(p,s)=0$,  the $s$ is the polarization or spin index of the spinor, and should  be distinguished from the $s$ quark or the energy $s$.
On the other hand, the current $J(0)$  also couples potentially to the $J^P={\frac{1}{2}}^+$   triply-charmed  pentaquark state $P^{+}$, as multiplying $i \gamma_{5}$ to the current $J(0)$ changes its parity  \cite{WangPc,Chung82,Bagan93,Oka96,WangHbaryon},
 \begin{eqnarray}
\langle 0| J (0)|P^{+}(p)\rangle &=&\lambda_{+} i\gamma_5 U^{+}(p,s) \, ,
\end{eqnarray}
the spinors $U^{\pm}(p,s)$    have analogous  properties.

On the phenomenological side,   we  insert  a complete set  of intermediate pentaquark states with the same quantum numbers as the current operators $J(x)$,
  and $i\gamma_5 J(x)$ into the correlation function $\Pi(p^2)$   to obtain the hadronic representation
\cite{SVZ79,Reinders85}. After isolating the pole terms of the lowest states of the triply-charmed  pentaquark states, we obtain the result:
\begin{eqnarray}
  \Pi(p^2) & = & \lambda_{-}^2 \,\, {\!\not\!{p}+ M_{-} \over M_{-}^{2}-p^{2}  }   +\lambda_{+}^2 \,\, {\!\not\!{p}- M_{+} \over M_{+}^{2}-p^{2}  }    +\cdots \, .
    \end{eqnarray}

Now we obtain the hadronic spectral density through dispersion relation,
\begin{eqnarray}
\frac{{\rm Im}\Pi(s)}{\pi}&=&\!\not\!{p} \left[\lambda^{2}_{-} \,\,\delta\left(s-M_{-}^2\right)+\lambda_{+}^2\,\, \delta\left(s-M_{+}^2\right)\right]+\left[M_{-}\lambda^{2}_{-} \,\,\delta\left(s-M_{-}^2\right)-M_{+}\lambda^{2}_{+} \,\,\delta\left(s-M_{+}^2\right)\right] \nonumber\\
&=&\!\not\!{p} \,\rho_H^1(s)+\rho_H^0(s) \, ,
\end{eqnarray}
where the subscript index $H$  denotes the hadron side,
then we introduce the weight function $\exp\left(-\frac{s}{T^2}\right)$ to obtain the QCD sum rules at the hadron  side,
\begin{eqnarray}
\int_{9m_c^2}^{s_0}ds \left[\sqrt{s}\rho_H^1(s)+\rho^0_{H}(s)\right]\exp\left( -\frac{s}{T^2}\right)
&=&2M_{-}\lambda^{2}_{-}\,\,\exp\left( -\frac{M_{-}^2}{T^2}\right) \, ,\\
\int_{9m_c^2}^{s_0}ds \left[\sqrt{s}\rho^1_{H}(s)-\rho^0_{H}(s)\right]\exp\left( -\frac{s}{T^2}\right)
&=&2M_{+}\lambda^{2}_{+}\,\,\exp\left( -\frac{M_{+}^2}{T^2}\right) \, ,
\end{eqnarray}
where the $s_0$ are the continuum threshold parameters and the $T^2$ are the Borel parameters. We separate the contributions of   the
negative-parity (positive-parity) pentaquark states from the positive-parity (negative-parity) pentaquark states explicitly. There is no contamination comes from the
positive or negative parity triply-charmed  pentaquark state.

In the following, we briefly outline  the operator product expansion for the correlation function $\Pi(p^2)$   in perturbative QCD. Firstly,  we contract the $u$, $d$ and $c$ quark fields in the correlation function $\Pi(p^2)$    with Wick theorem, and obtain the result:
\begin{eqnarray}
\Pi(p^2)&=&-i\,\varepsilon^{ila}\varepsilon^{ijk}\varepsilon^{lmn}\varepsilon^{i^{\prime}l^{\prime}a^{\prime}}\varepsilon^{i^{\prime}j^{\prime}k^{\prime}}
\varepsilon^{l^{\prime}m^{\prime}n^{\prime}}\int d^4x\, e^{ip\cdot x} \nonumber\\
&&\left\{  {\rm Tr}\left[\gamma_5 C U^{T}_{jj^\prime}(x)C  \gamma_5 C_{kk^\prime}(x)\right] \,{\rm Tr}\left[\gamma_5 C D^{T}_{mm^\prime}(x)C  \gamma_5 C_{nn^\prime}(x)\right] C C_{a^{\prime}a}^T(-x)C   \right. \nonumber\\
&&\left.- {\rm Tr} \left[\gamma_5 CU_{jj^\prime}^T(x)C \gamma_5  C_{kn^\prime}(x) \gamma_5 C D_{mm^\prime}^T(x)C \gamma_5 C_{nk^\prime}(x)\right]    C C_{a^{\prime}a}^T(-x)C \right\} \, ,
\end{eqnarray}
where
the $U_{ij}(x)$, $D_{ij}(x)$ and $C_{ij}(x)$ are the full $u$, $d$ and $c$ quark propagators, respectively (we can set $S_{ij}(x)=U_{ij}(x),\,D_{ij}(x)$ in the chiral limit $m_u=m_d=0$),
 \begin{eqnarray}
S_{ij}(x)&=& \frac{i\delta_{ij}\!\not\!{x}}{ 2\pi^2x^4}-\frac{\delta_{ij}\langle
\bar{q}q\rangle}{12} -\frac{\delta_{ij}x^2\langle \bar{q}g_s\sigma Gq\rangle}{192} -\frac{ig_sG^{a}_{\alpha\beta}t^a_{ij}(\!\not\!{x}
\sigma^{\alpha\beta}+\sigma^{\alpha\beta} \!\not\!{x})}{32\pi^2x^2}  \nonumber\\
&&  -\frac{1}{8}\langle\bar{q}_j\sigma^{\mu\nu}q_i \rangle \sigma_{\mu\nu}+\cdots \, ,
\end{eqnarray}
\begin{eqnarray}
C_{ij}(x)&=&\frac{i}{(2\pi)^4}\int d^4k e^{-ik \cdot x} \left\{
\frac{\delta_{ij}}{\!\not\!{k}-m_c}
-\frac{g_sG^n_{\alpha\beta}t^n_{ij}}{4}\frac{\sigma^{\alpha\beta}(\!\not\!{k}+m_c)+(\!\not\!{k}+m_c)
\sigma^{\alpha\beta}}{(k^2-m_c^2)^2}\right.\nonumber\\
&&\left. -\frac{g_s^2 (t^at^b)_{ij} G^a_{\alpha\beta}G^b_{\mu\nu}(f^{\alpha\beta\mu\nu}+f^{\alpha\mu\beta\nu}+f^{\alpha\mu\nu\beta}) }{4(k^2-m_c^2)^5}+\cdots\right\} \, ,\nonumber\\
f^{\alpha\beta\mu\nu}&=&(\!\not\!{k}+m_c)\gamma^\alpha(\!\not\!{k}+m_c)\gamma^\beta(\!\not\!{k}+m_c)\gamma^\mu(\!\not\!{k}+m_c)\gamma^\nu(\!\not\!{k}+m_c)\, ,
\end{eqnarray}
and  $t^n=\frac{\lambda^n}{2}$, the $\lambda^n$ is the Gell-Mann matrix
\cite{Reinders85,WangHuang-3900,Pascual-1984}.
We retain the term $\langle\bar{q}_j\sigma_{\mu\nu}q_i \rangle$  originates from  Fierz re-ordering of the $\langle q_i \bar{q}_j\rangle$ to  absorb the gluons  emitted from other quark lines to form $\langle\bar{q}_j g_s G^a_{\alpha\beta} t^a_{mn}\sigma_{\mu\nu} q_i \rangle$   to extract the mixed condensate   $\langle\bar{q}g_s\sigma G q\rangle$ \cite{WangHuang-3900}.
Then we compute  the integrals both in the coordinate and momentum spaces to obtain the correlation function $\Pi(p^2)$,   therefore the QCD spectral densities $\rho_{QCD}^1(s)$ and $\rho_{QCD}^0(s)$ at the quark level through dispersion  relation,
 \begin{eqnarray}
\frac{{\rm Im}\Pi(s)}{\pi}&=&\!\not\!{p} \,\rho^1_{QCD}(s)+m_c\rho^0_{QCD}(s) \, .
\end{eqnarray}
  In this article, we carry out the
operator product expansion to the vacuum condensates  up to dimension-10, and
assume vacuum saturation for the  higher dimensional vacuum condensates. The condensates $\langle g_s^3 GGG\rangle$, $\langle \frac{\alpha_s GG}{\pi}\rangle^2$,
 $\langle \frac{\alpha_s GG}{\pi}\rangle\langle \bar{q} g_s \sigma Gq\rangle$ have the dimensions 6, 8, 9 respectively,  but they are   the vacuum expectations
of the operators of the order    $\mathcal{O}( \alpha_s^{3/2})$, $\mathcal{O}(\alpha_s^2)$, $\mathcal{O}( \alpha_s^{3/2})$ respectively, and discarded \cite{WangHuang-3900}.

 Once the analytical QCD spectral densities $\rho_{QCD}^1(s)$ and $\rho_{QCD}^0(s)$ are obtained,  we can take the
quark-hadron duality below the continuum thresholds  $s_0$ and introduce the weight function $\exp\left(-\frac{s}{T^2}\right)$ to obtain  the   QCD sum rules:
\begin{eqnarray}
2M_{-}\lambda^{2}_{-}\,\,\exp\left( -\frac{M_{-}^2}{T^2}\right)
&=& \int_{9m_c^2}^{s_0}ds \left[\sqrt{s}\rho_{QCD}^1(s)+m_c\rho_{QCD}^0(s)\right]\exp\left( -\frac{s}{T^2}\right)\, ,
\end{eqnarray}
\begin{eqnarray}
2M_{+}\lambda^{2}_{+}\,\,\exp\left( -\frac{M_{+}^2}{T^2}\right)
&=& \int_{9m_c^2}^{s_0}ds \left[\sqrt{s}\rho_{QCD}^1(s)-m_c\rho_{QCD}^0(s)\right]\exp\left( -\frac{s}{T^2}\right)\, ,
\end{eqnarray}
where

\begin{eqnarray}
\rho_{QCD}^1(s)&=&\rho_{0}^1(s)+\rho_{3}^1(s)+\rho_{4}^1(s)+\rho_{5}^1(s)+\rho_{6}^1(s)+\rho_{8}^1(s)+\rho_{10}^1(s)\, , \nonumber\\
\rho_{QCD}^0(s)&=&\rho_{0}^0(s)+\rho_{3}^0(s)+\rho_{4}^0(s)+\rho_{5}^0(s)+\rho_{6}^0(s)+\rho_{8}^0(s)+\rho_{10}^0(s)\, ,
\end{eqnarray}
\begin{eqnarray}
\rho_{0}^1(s)&=&\frac{1}{40960\pi^8}\int dztr \, ztr\,(1-r-t-z)^2 \left(s-\widehat{m}_c^2 \right)^4\left(8s-3\widehat{m}_c^2 \right)\, ,\nonumber\\
\rho_{0}^0(s)&=&\frac{1}{40960\pi^8}\int dztr \, zt\,(1-r-t-z)^2 \left(s-\widehat{m}_c^2 \right)^4\left(7s-2\widehat{m}_c^2 \right) \, ,
\end{eqnarray}

\begin{eqnarray}
\rho_{3}^1(s)&=&-\frac{m_c\langle\bar{q}q\rangle}{128\pi^6}\int dztr \, rz\,(1-r-t-z) \left(s-\widehat{m}_c^2 \right)^2\left(2s-\widehat{m}_c^2 \right)\, ,\nonumber\\
\rho_{3}^0(s)&=&-\frac{m_c\langle\bar{q}q\rangle}{384\pi^6}\int dztr \, z\,(1-r-t-z) \left(s-\widehat{m}_c^2 \right)^2\left(5s-2\widehat{m}_c^2 \right) \, ,
\end{eqnarray}

\begin{eqnarray}
\rho_{4}^1(s)&=&-\frac{m_c^2 }{6144\pi^6}\langle\frac{\alpha_sGG}{\pi}\rangle\int dztr \,t\, \left(\frac{r}{z^2}+\frac{z}{2r^2} \right)(1-r-t-z)^2\left(s-\widehat{m}_c^2 \right)\left(5s-3\widehat{m}_c^2 \right) \nonumber\\
&&+\frac{1}{2048\pi^6}\langle\frac{\alpha_sGG}{\pi}\rangle\int dztr \, rt(1-r-t-z) \left(s-\widehat{m}_c^2 \right)^2\left(2s-\widehat{m}_c^2 \right) \nonumber\\
&&+\frac{3}{65536\pi^6}\langle\frac{\alpha_sGG}{\pi}\rangle\int dztr \, rtz \left(s-\widehat{m}_c^2 \right)^2\left(2s-\widehat{m}_c^2 \right) \nonumber\\
&&+\frac{3m_c^2}{65536\pi^6}\langle\frac{\alpha_sGG}{\pi}\rangle\int dztr \, r  \left(s-\widehat{m}_c^2 \right)^2  \nonumber\\
&&-\frac{3}{131072\pi^6}\langle\frac{\alpha_sGG}{\pi}\rangle\int dztr \, r(1-r-t-z)^2 \left(s-\widehat{m}_c^2 \right)^2\left(2s-\widehat{m}_c^2 \right)\, , \nonumber\\
\rho_{4}^0(s)&=&-\frac{m_c^2 }{3072\pi^6}\langle\frac{\alpha_sGG}{\pi}\rangle\int dztr \, t\, \left(\frac{1}{z^2}+\frac{z}{2r^3} \right)(1-r-t-z)^2\left(s-\widehat{m}_c^2 \right)\left(2s-\widehat{m}_c^2 \right) \nonumber\\
&&+\frac{1 }{12288\pi^6}\langle\frac{\alpha_sGG}{\pi}\rangle\int dztr \,\left( \frac{tz}{r^2}-\frac{3}{32} \right) (1-r-t-z)^2\left(s-\widehat{m}_c^2 \right)^2\left(5s-2\widehat{m}_c^2 \right) \nonumber\\
&&+\frac{1}{6144\pi^6}\langle\frac{\alpha_sGG}{\pi}\rangle\int dztr \, t(1-r-t-z) \left(s-\widehat{m}_c^2 \right)^2\left(5s-2\widehat{m}_c^2 \right) \nonumber\\
&&+\frac{1}{65536\pi^6}\langle\frac{\alpha_sGG}{\pi}\rangle\int dztr \, tz \left(s-\widehat{m}_c^2 \right)^2\left(5s-2\widehat{m}_c^2 \right) \nonumber\\
&&+\frac{3m_c^2}{65536\pi^6}\langle\frac{\alpha_sGG}{\pi}\rangle\int dztr \,   \left(s-\widehat{m}_c^2 \right)^2 \, ,
\end{eqnarray}

\begin{eqnarray}
\rho_{5}^1(s)&=&\frac{m_c\langle\bar{q}g_s\sigma Gq\rangle}{512\pi^6}\int dztr \, rz\,\left(1- \frac{1-r-t-z}{t}-\frac{3}{16}\right) \left(s-\widehat{m}_c^2 \right)\left(5s-3\widehat{m}_c^2 \right) \, ,\nonumber\\
\rho_{5}^0(s)&=&\frac{m_c\langle\bar{q}g_s\sigma Gq\rangle}{256\pi^6}\int dztr \, z\,\,\left(1- \frac{1-r-t-z}{t}-\frac{3}{16}\right) \left(s-\widehat{m}_c^2 \right)\left(2s-\widehat{m}_c^2 \right) \, ,
\end{eqnarray}

\begin{eqnarray}
\rho_{6}^1(s)&=&\frac{m_c^2\langle\bar{q}q\rangle^2}{96\pi^4}\int dzt \, (1-t-z) \left(s-\overline{m}_c^2 \right)\, ,\nonumber\\
\rho_{6}^0(s)&=&\frac{m_c^2\langle\bar{q}q\rangle^2}{96\pi^4}\int dzt \,  \left(s-\overline{m}_c^2 \right)\, ,
\end{eqnarray}

\begin{eqnarray}
\rho_{8}^1(s)&=&-\frac{m_c^2\langle\bar{q}q\rangle \langle\bar{q}g_s\sigma Gq\rangle}{64\pi^4}\int dzt \, (1-t-z)\left[1+\frac{s}{3}\delta \left(s-\overline{m}_c^2 \right)\right] \nonumber\\
&&+\frac{m_c^2\langle\bar{q}q\rangle \langle\bar{q}g_s\sigma Gq\rangle}{192\pi^4}\int dzt \, \frac{1-t-z}{t }\, , \nonumber\\
\rho_{8}^0(s)&=&-\frac{m_c^2\langle\bar{q}q\rangle \langle\bar{q}g_s\sigma Gq\rangle}{96\pi^4}\int dzt \, \left[1+\frac{s}{2}\delta \left(s-\overline{m}_c^2 \right)\right] \nonumber\\
&&+\frac{m_c^2\langle\bar{q}q\rangle \langle\bar{q}g_s\sigma Gq\rangle}{192\pi^4}\int dzt \, \frac{1}{t } \, ,
\end{eqnarray}

\begin{eqnarray}
\rho_{10}^1(s)&=&\frac{m_c^2 \langle\bar{q}g_s\sigma Gq\rangle^2}{256\pi^4}\int dzt \, (1-t-z)\left(1+\frac{2s}{3T^2}+\frac{s^2}{6T^4}\right)\delta \left(s-\overline{m}_c^2 \right) \nonumber\\
&&-\frac{m_c^2 \langle\bar{q}g_s\sigma Gq\rangle^2}{384\pi^4}\int dzt \, \frac{1-t-z}{t}\left(1+\frac{s}{2T^2}\right)\delta \left(s-\overline{m}_c^2 \right) \nonumber\\
&&+\frac{11m_c^2 \langle\bar{q}g_s\sigma Gq\rangle^2}{18432\pi^4}\int dzt \, \frac{1-t-z}{tz} \delta \left(s-\overline{m}_c^2 \right)\, , \nonumber\\
\rho_{10}^0(s)&=&\frac{m_c^2 \langle\bar{q}g_s\sigma Gq\rangle^2}{768\pi^4}\int dzt \,  \left(1+\frac{s}{T^2}+\frac{s^2}{2T^4}\right)\delta \left(s-\overline{m}_c^2 \right) \nonumber\\
&&-\frac{m_c^2 \langle\bar{q}g_s\sigma Gq\rangle^2}{768\pi^4}\int dzt \, \frac{1}{t}\left(1+\frac{s}{T^2}\right)\delta \left(s-\overline{m}_c^2 \right) \nonumber\\
&&+\frac{11m_c^2 \langle\bar{q}g_s\sigma Gq\rangle^2}{18432\pi^4}\int dzt \, \frac{1}{tz} \delta \left(s-\overline{m}_c^2 \right) \, ,
\end{eqnarray}

\begin{eqnarray}
\int dztr &=& \int_{z_i}^{z_f} dz \int_{t_i}^{t_f}dt \int_{r_i}^{r_f}dr \, ,\nonumber\\
\int dzt &=& \int_{z_i}^{z_f} dz \int_{t_i}^{t_f}dt  \, ,
\end{eqnarray}

\begin{eqnarray}
z_{f/i}&=&\frac{\hat{s}-3 \pm \sqrt{(\hat{s}-3)^2-4\hat{s}}}{2\hat{s}} \, ,\nonumber\\
t_{f/i}&=&\frac{1-z \pm \sqrt{(1-z)^2-4\frac{z-z^2}{z\hat{s}-1}}}{2 } \, ,\nonumber\\
r_i&=&\frac{tz}{tz\hat{s}-t-z}\, , \nonumber\\
r_f&=&1-z-t\, ,
\end{eqnarray}

\begin{eqnarray}
\widehat{m}_c^2&=&\frac{m_c^2}{t}+\frac{m_c^2}{z}+\frac{m_c^2}{r}\, ,\nonumber\\
\overline{m}_c^2&=&\frac{m_c^2}{t}+\frac{m_c^2}{z}+\frac{m_c^2}{1-t-z}\, , \nonumber\\
\hat{s}&=&\frac{
s}{m_c^2}\, ,
\end{eqnarray}
\begin{eqnarray}
\int dzt &\to& \int_{0}^{1} dz \int_{0}^{1-z}dt  \, ,
\end{eqnarray}
  when the $\delta$ function $\delta\left(s-\overline{m}_c^2\right)$  appears.

We derive    Eqs.(14-15) with respect to  $\tau=\frac{1}{T^2}$, then eliminate the
 pole residues $\lambda_{\pm}$ and obtain the QCD sum rules for
 the masses of the triply-charmed pentaquark states,
 \begin{eqnarray}
 M^2_{-} &=& \frac{-\frac{d}{d\tau}\int_{9m_c^2}^{s_0}ds \left[\sqrt{s}\rho_{QCD}^1(s)+m_c\rho_{QCD}^0(s)\right]\exp\left( -s\tau\right)}{\int_{9m_c^2}^{s_0}ds \left[\sqrt{s}\rho_{QCD}^1(s)+m_c\rho_{QCD}^0(s)\right]\exp\left( -s\tau\right)}\, ,\\
 M^2_{+} &=& \frac{-\frac{d}{d\tau}\int_{9m_c^2}^{s_0}ds \left[\sqrt{s}\rho_{QCD}^1(s)-m_c\rho_{QCD}^0(s)\right]\exp\left( -s\tau\right)}{\int_{9m_c^2}^{s_0}ds \left[\sqrt{s}\rho_{QCD}^1(s)-m_c\rho_{QCD}^0(s)\right]\exp\left( -s\tau\right)}\, .
\end{eqnarray}

\section{Numerical results and discussions}

We take  the standard values of the vacuum condensates $\langle
\bar{q}q \rangle=-(0.24\pm 0.01\, \rm{GeV})^3$,   $\langle
\bar{q}g_s\sigma G q \rangle=m_0^2\langle \bar{q}q \rangle$,
$m_0^2=(0.8 \pm 0.1)\,\rm{GeV}^2$,  $\langle \frac{\alpha_s
GG}{\pi}\rangle=(0.33\,\rm{GeV})^4 $    at the energy scale  $\mu=1\, \rm{GeV}$
\cite{SVZ79,Reinders85,Colangelo-Review}, and take the $\overline{MS}$ mass  $m_{c}(m_c)=(1.28\pm0.03)\,\rm{GeV}$
 from the Particle Data Group \cite{PDG}. Furthermore, we set $m_u=m_d=0$ due to the small current quark masses.
 We take into account
the energy-scale dependence of  the input parameters from the renormalization group equation,
\begin{eqnarray}
\langle\bar{q}q \rangle(\mu)&=&\langle\bar{q}q \rangle(Q)\left[\frac{\alpha_{s}(Q)}{\alpha_{s}(\mu)}\right]^{\frac{12}{25}}\, ,\nonumber\\
\langle\bar{q}g_s \sigma Gq \rangle(\mu)&=&\langle\bar{q}g_s \sigma Gq \rangle(Q)\left[\frac{\alpha_{s}(Q)}{\alpha_{s}(\mu)}\right]^{\frac{2}{25}}\, , \nonumber\\
m_c(\mu)&=&m_c(m_c)\left[\frac{\alpha_{s}(\mu)}{\alpha_{s}(m_c)}\right]^{\frac{12}{25}} \, ,\nonumber\\
\alpha_s(\mu)&=&\frac{1}{b_0t}\left[1-\frac{b_1}{b_0^2}\frac{\log t}{t} +\frac{b_1^2(\log^2{t}-\log{t}-1)+b_0b_2}{b_0^4t^2}\right]\, ,
\end{eqnarray}
  where $t=\log \frac{\mu^2}{\Lambda^2}$, $b_0=\frac{33-2n_f}{12\pi}$, $b_1=\frac{153-19n_f}{24\pi^2}$, $b_2=\frac{2857-\frac{5033}{9}n_f+\frac{325}{27}n_f^2}{128\pi^3}$,  $\Lambda=210\,\rm{MeV}$, $292\,\rm{MeV}$  and  $332\,\rm{MeV}$ for the flavors  $n_f=5$, $4$ and $3$, respectively  \cite{PDG,Narison-mix,Narison-Book}, and evolve all the input parameters to the optimal energy scales   to extract the  masses of the triply-charmed pentaquark states with the flavor $n_f=4$.

In the article, we study the scalar-diquark-scalar-diquark-antiquark type  pentaquark state,
which consists of   two charmed diquark states and a charmed antiquark.
  In the heavy quark limit, the $c$-quark serves as a static well potential and  combines with a  light quark $q$  to form a charmed  diquark  in  color antitriplet,
or combines with a  light antiquark $\bar{q}$ to form a charmed  meson in color singlet (meson-like state in color octet),
\begin{eqnarray}
q^j+c^k &\to & \varepsilon^{ijk}\, q^j\,c^k\, , \nonumber\\
\bar{q}^{ j}+c^k &\to & \bar{q}^{ j} \,\delta_{jk}\, c^k\,\, (\bar{q}^{ j}\,\lambda^{a}_{jk}\,c^k) \, ,
\end{eqnarray}
where the $i$, $j$, $k$ are color indexes, the $\lambda^a$ is Gell-Mann matrix.
 Then
\begin{eqnarray}
 \varepsilon^{ijk}\, q^j\,c^k+\varepsilon^{lmn} q^{\prime m}\,c^n+\bar{c}^a &\to &  {\rm compact \,\,\, pentaquark \,\,\, states}\,  .
\end{eqnarray}

The five-quark systems $qq^\prime cc\bar{c}$   are characterized by the effective charmed quark mass ${\mathbb{M}}_c$ (or constituent quark mass) and the virtuality  $V=\sqrt{M^2_{P}-(3{\mathbb{M}}_c)^2}$ (or bound energy not as robust), where the   $P$ denotes  the triply-charmed pentaquark  states. It is natural to set the energy  scales of the QCD spectral densities to be $\mu=V$.
In Refs.\cite{WangHuang-3900,Wang-4430,WangMolecule-3900,Wang-4660-2014,WangPc,Wang-HexaQuark,Wang-cc-tetraquark,Wang-4025-CTP,WangHuang-NPA-2014}, we study the acceptable energy scales of the QCD spectral densities  for the hidden-charm (hidden-bottom) tetraquark states and molecular states, hidden-charm pentaquark states, hidden-charm hexaquark states, and  doubly-heavy tetraquark states in the QCD sum rules in details,  and suggest an energy scale formula  $\mu=\sqrt{M^2_{X/Y/Z/P}-(2{\mathbb{M}}_Q)^2}$ to determine  the optimal   energy scales, which works well. The updated values of the effective heavy quark masses are ${\mathbb{M}}_c=1.82\,\rm{GeV}$ and ${\mathbb{M}}_b=5.17\,\rm{GeV}$ for the multiquark states having heavy-light diquark states  \cite{Wang-1601}.
Now we use the energy scale formula,
\begin{eqnarray}
\mu&=&\sqrt{M^2_{P}-(3{\mathbb{M}}_c)^2}\, ,
\end{eqnarray}
 to determine the ideal energy scales of the QCD spectral densities.

In this article, we take the continuum threshold parameters  as  $\sqrt{s_0}=M_{P}+(0.4\sim0.7)\,\rm{GeV}$, and vary the parameters $\sqrt{s_0}$ to obtain   the optimal Borel parameters $T^2$ to satisfy  the  following four criteria:

$\bf 1.$ Pole dominance on the phenomenological side;

$\bf 2.$ Convergence of the operator product expansion;

$\bf 3.$ Appearance of the Borel platforms;

$\bf 4.$ Satisfying the energy scale formula.

In calculations, we observe that
\begin{eqnarray}
\mu\uparrow  \, \, \, \, \,  M_P \downarrow \, ,\nonumber\\
\mu\downarrow  \, \, \, \, \,  M_P \uparrow \, ,
\end{eqnarray}
from the QCD sum rules  in Eqs.(28-29). On the other hand, the energy scale formula  indicates that
\begin{eqnarray}
\mu\uparrow  \, \, \, \, \,  M_P \uparrow \, ,\nonumber\\
\mu\downarrow  \, \, \, \, \,  M_P \downarrow \, ,
\end{eqnarray}
as it can be rewritten as
\begin{eqnarray}
M_P&=& \sqrt{\mu^2+(3{\mathbb{M}}_c)^2}\, .
\end{eqnarray}
It is difficult to obtain the optimal energy scales  $\mu$ and masses $M_P$, however, the optimal energy scales  $\mu$ and masses $M_P$ do exist.
The resulting  Borel parameters or Borel windows $T^2$, continuum threshold parameters $s_0$, optimal energy scales of the QCD spectral densities, pole contributions of the ground states are shown explicitly in Table 1.

In Fig.1, we plot the contributions of the vacuum condensates $D_n$ of dimension $n$ in the operator product expansion  for the central values of the input parameters,
   \begin{eqnarray}
D_n&=& \frac{  \int_{9m_c^2}^{s_0} ds\,\left[\sqrt{s}\rho_{n}^1(s)\pm m_c\rho_{n}^0(s)\right]\,\exp\left(-\frac{s}{T^2}\right)}{\int_{9m_c^2}^{s_0} ds \,\left[\sqrt{s}\rho_{QCD}^1(s)\pm m_c\rho_{QCD}^0(s)\right]\,\exp\left(-\frac{s}{T^2}\right)}\, .
\end{eqnarray}
    From the figure, we can see that the dominant contributions come from the quark condensate $D_3$, the contributions of the perturbative terms (or $D_0$) are about $(20-30)\%$, so in this article we approximate the continuum contributions as $\left[\sqrt{s}\rho_{QCD}^1(s)\pm m_c\rho_{QCD}^0(s)\right]\Theta(s-s_0)$, and define
    the pole contributions $\rm{PC}$ as
   \begin{eqnarray}
{\rm PC}&=& \frac{  \int_{9m_c^2}^{s_0} ds\,\left[\sqrt{s}\rho_{QCD}^1(s)\pm  m_c\rho_{QCD}^0(s)\right]\,\exp\left(-\frac{s}{T^2}\right)}{\int_{9m_c^2}^{\infty} ds \,\left[\sqrt{s}\rho_{QCD}^1(s)\pm m_c\rho_{QCD}^0(s)\right]\,\exp\left(-\frac{s}{T^2}\right)}\, .
\end{eqnarray}
 From Table 1, we can see that the pole dominance condition can be well satisfied.
   Although the contributions of the vacuum condensate of dimension $n=3$ are very large,  the contributions of the vacuum condensates of dimensions $6, \,8,\,10$ have the hierarchy $D_6\gg |D_8|\gg |D_{10}|$, the operator product expansion is  convergent. Now the criterion ${\bf 1}$ and criterion ${\bf 2}$ are satisfied.

\begin{figure}
 \centering
 \includegraphics[totalheight=5cm,width=7cm]{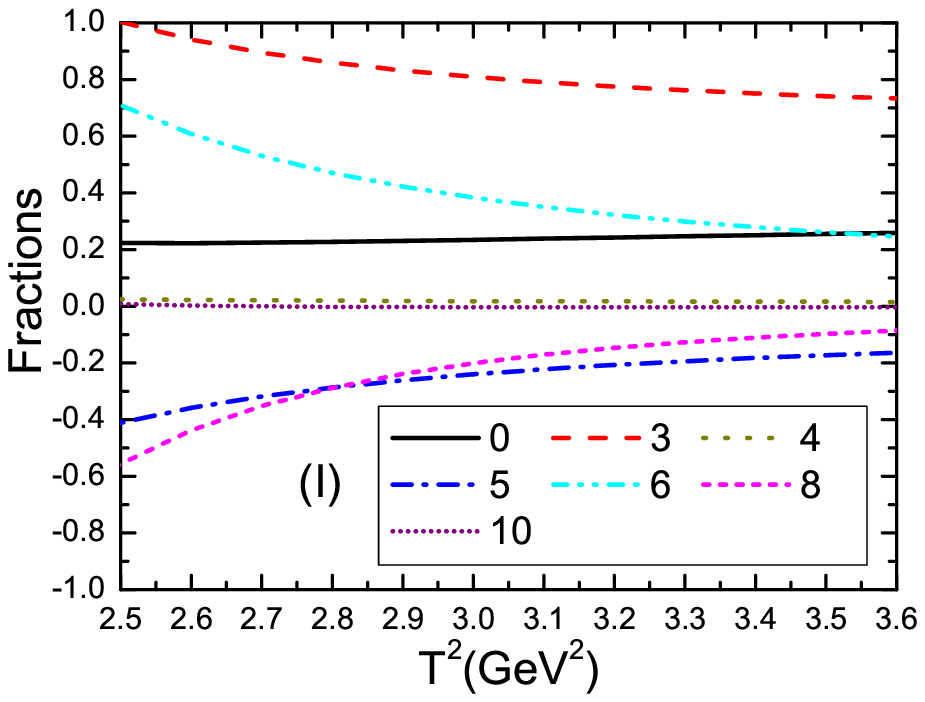}
 \includegraphics[totalheight=5cm,width=7cm]{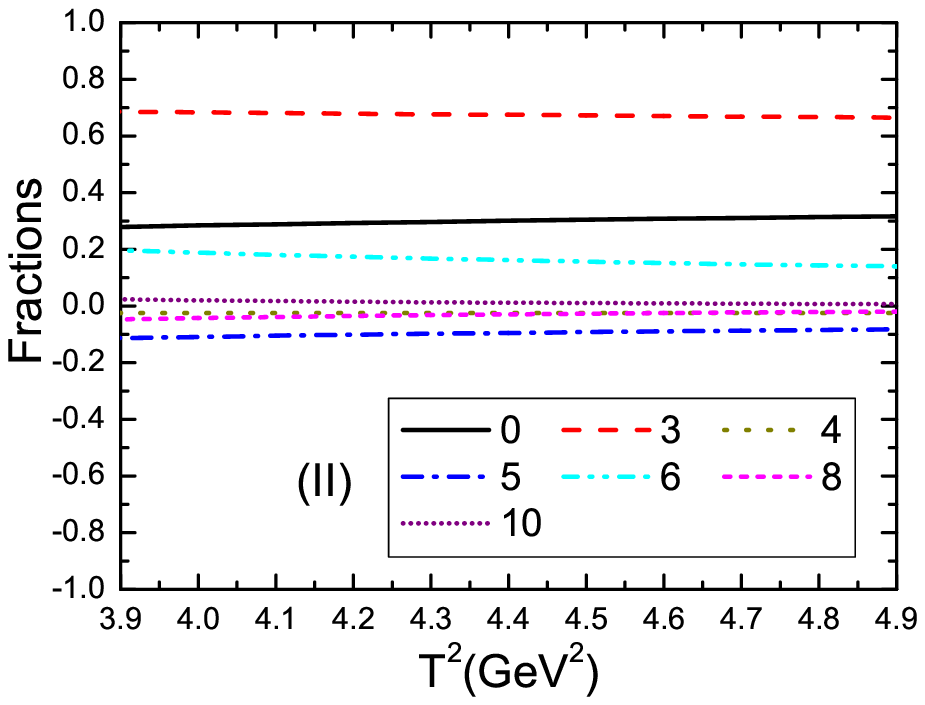}
          \caption{ The contributions of the vacuum condensations of dimension $n$ with $n=0$, $3$, $4$, $5$, $6$, $8$ and $10$, where  the (I) and (II) denote the negative parity and positive parity  pentaquark states, respectively.   }
\end{figure}

We take  into account all uncertainties of the input parameters,
and obtain the values of the masses and pole residues of
 the   triply-charmed pentaquark states, which are  shown explicitly in Table 1 and Figs.2-3.
 In Figs.2-3, we plot the masses and pole residues of the triply-charmed pentaquark states in  much larger  ranges than the Borel windows.
 From the figures, we can see that the platforms for the mass and pole residue of the $J^P={\frac{1}{2}}^-$ pentaquark state appear as the minimum values,   the Borel platforms are very flat, while the predicted mass and pole residue of the $J^P={\frac{1}{2}}^+$ pentaquark state increase slowly with the increase of the Borel parameter, we determine the platform  by requiring the uncertainty $\frac{\delta M_{P}}{M_{P}} $ induced by the Borel parameter is less than $1\%$. The criterion ${\bf 3}$ is also satisfied, furthermore, the energy scale formula $\mu=\sqrt{M^2_{P}-(3{\mathbb{M}}_c)^2}$ is well satisfied. Now the four criteria are all satisfied, we expect to make reliable predictions.

\begin{table}
\begin{center}
\begin{tabular}{|c|c|c|c|c|c|c|c|}\hline\hline
                    &$T^2(\rm{GeV}^2)$   &$\sqrt{s_0}(\rm{GeV})$   &$\mu(\rm{GeV})$  &pole          &$M(\rm{GeV})$  &$\lambda(\rm{GeV}^5)$ \\ \hline

${\frac{1}{2}}^-$   &$2.9-3.3$           &$6.3\pm0.1$              &$1.3$            &$(71-87)\%$   &$5.61\pm0.10$  &$(2.38\pm0.31)\times10^{-3}$   \\ \hline

${\frac{1}{2}}^+$   &$4.1-4.5$           &$6.4\pm0.1$              &$1.7$            &$(42-61)\%$   &$5.72\pm0.10$  &$(1.45\pm0.28)\times10^{-3}$   \\ \hline\hline
\end{tabular}
\end{center}
\caption{ The Borel parameters (Borel windows), continuum threshold parameters, optimal  energy scales, pole contributions,   masses and pole residues for the
 triply-charmed   pentaquark states. }
\end{table}

\begin{figure}
 \centering
 \includegraphics[totalheight=5cm,width=7cm]{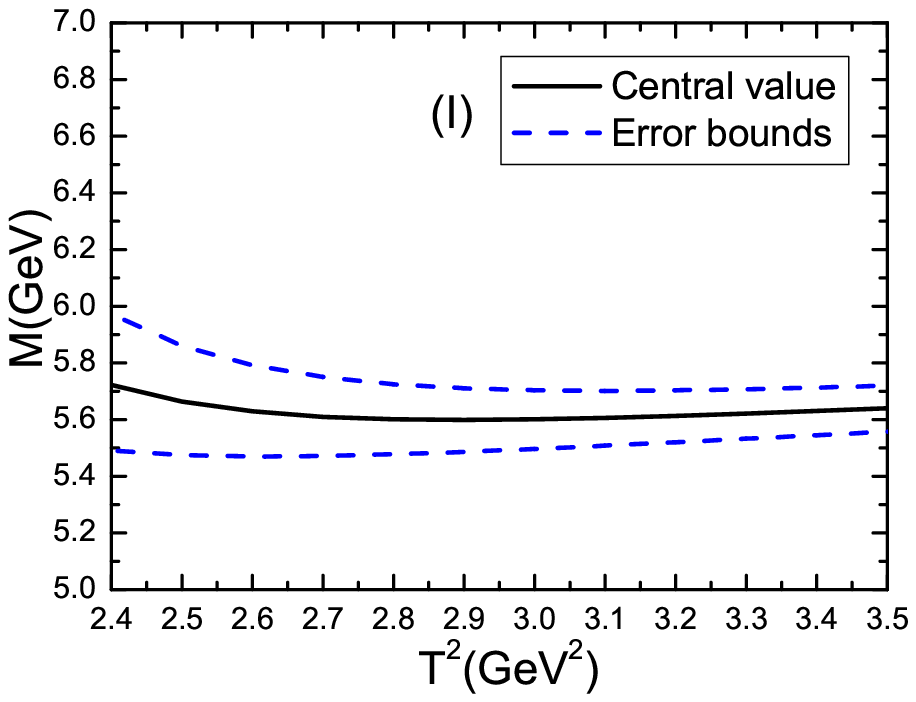}
 \includegraphics[totalheight=5cm,width=7cm]{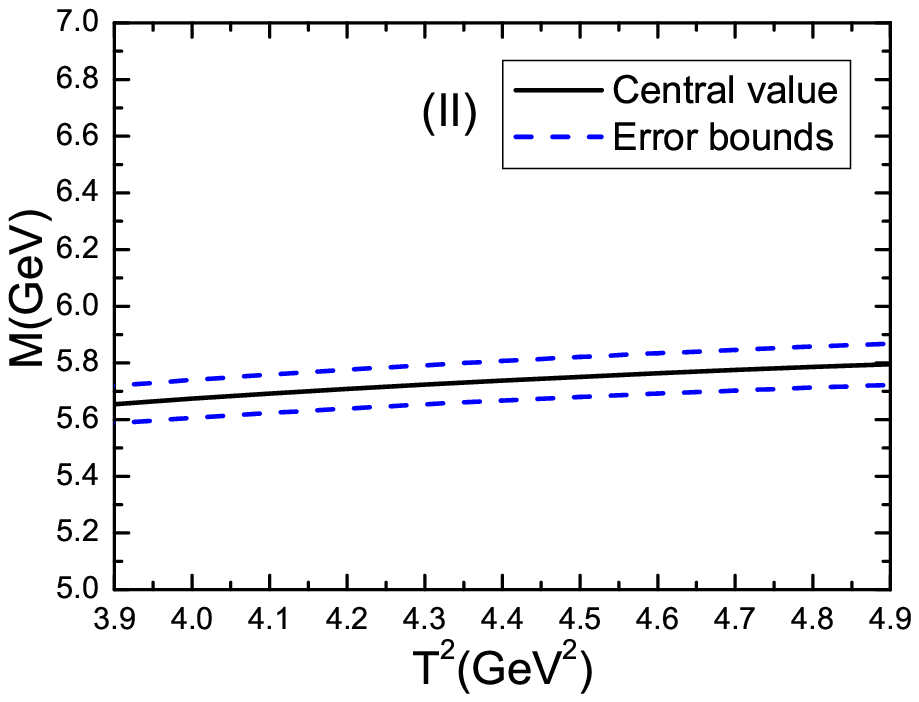}
         \caption{ The masses of the triply-charmed pentaquark states with variations of the Borel parameters, where  the (I) and (II) denote the negative parity and positive parity  pentaquark states, respectively.   }
\end{figure}

\begin{figure}
 \centering
 \includegraphics[totalheight=5cm,width=7cm]{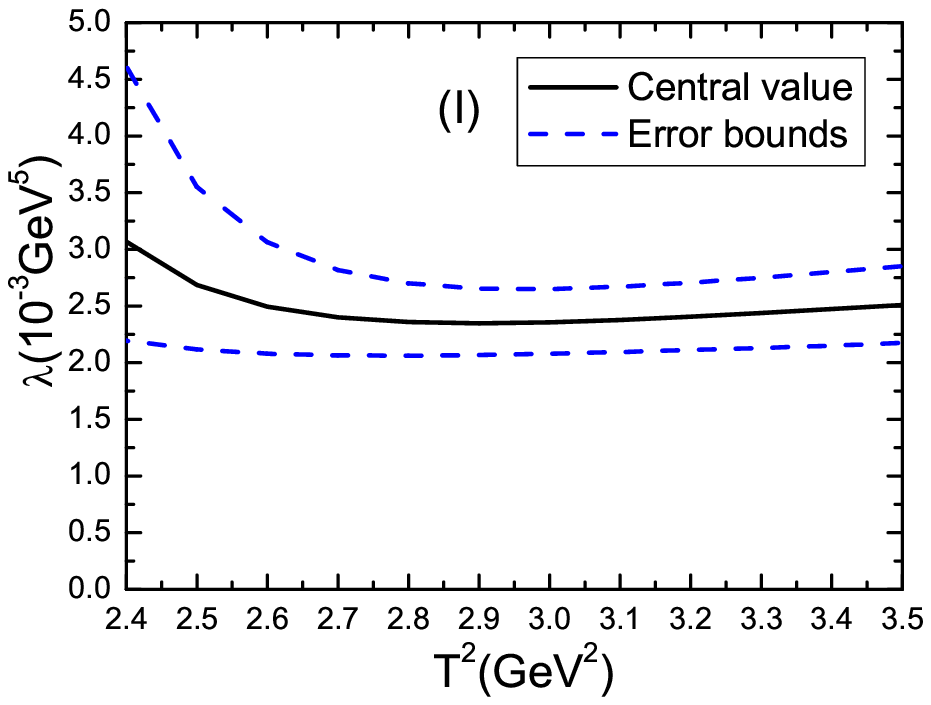}
 \includegraphics[totalheight=5cm,width=7cm]{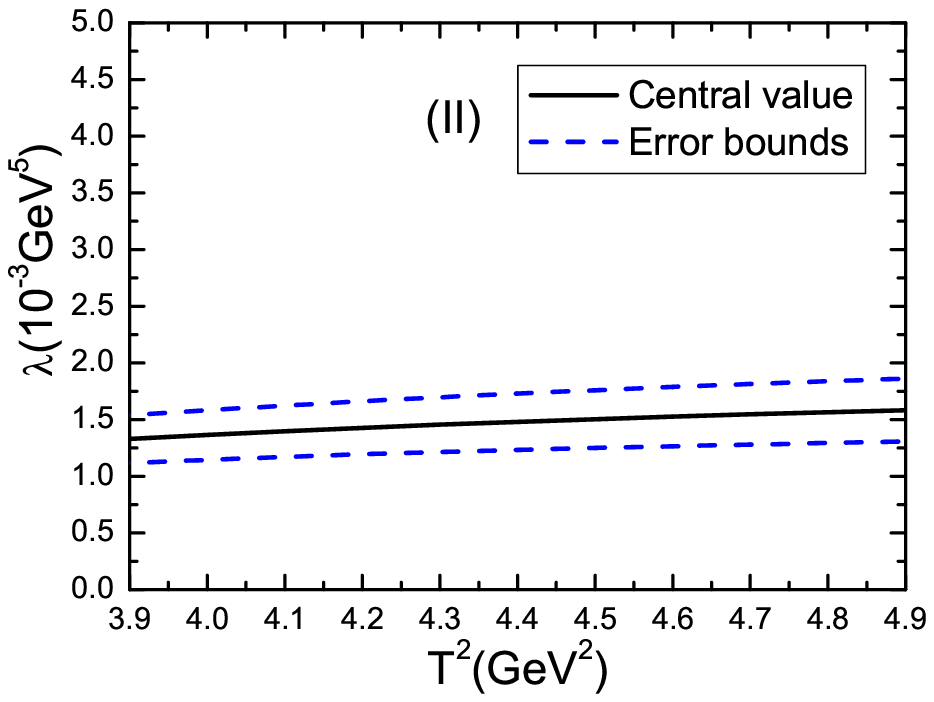}
          \caption{ The pole residues of the triply-charmed pentaquark states with variations of the Borel parameters, where  the (I) and (II) denote the negative parity and positive parity  pentaquark states, respectively.   }
\end{figure}

If the energy scale formula survives, the masses of the lowest triply-charmed pentaquark states should be larger than $\sqrt{{(\rm 1GeV)^2}+(3{\mathbb{M}}_c)^2}=5.55\,\rm{GeV}$.   In Fig.4, we plot the predicted masses of the negative parity and positive parity triply-charmed pentaquark states  with variations of the Borel parameters $T^2$ for  the threshold  parameters $\sqrt{s_0}=6.3\,\rm{GeV}$ and $6.4\,\rm{GeV}$, respectively. From the figure, we can see that the most flat platform appears at the energy scale $\mu=1.3\,\rm{GeV}$ for the $J^P={\frac{1}{2}}^-$  triply-charmed pentaquark state, which happens to be the optimal energy scale determined by the energy scale formula. For the $J^P={\frac{1}{2}}^+$  triply-charmed pentaquark state,  no platform is more flat than others, we determine the optimal energy scale by the energy scale formula $\mu=\sqrt{M^2_{P}-(3{\mathbb{M}}_c)^2}$. In this article, we obtain the continuum threshold parameters $\sqrt{s_0}=M_P+(0.6\sim 0.8)\,\rm{GeV}$.
In previous works \cite{WangPc,WangHbaryon}, we study  the heavy, doubly-heavy, triply heavy baryon states and hidden-charm pentaquark states in a systematic way with  the QCD sum rules, the continuum threshold parameters $\sqrt{s_0}=M_{gr}+(0.6\sim 0.8)\,\rm{GeV}$ work well, where the $gr$ denotes the ground states. We expect that  the relation survives for the triply-charmed pentaquark states.

 In the QCD sum rules for the $M_{+}$, we choose much larger Borel parameter $T^2$ than that for the $M_{-}$, see Table 1, the contributions of the higher dimensional vacuum condensates especially the terms associate with $\frac{1}{T^2}$ and $\frac{1}{T^4}$ in the QCD spectral density are greatly suppressed. Moreover, in the QCD sum rules for the $M_{+}$,
the higher dimensional vacuum condensates obtain additional suppression due to the special combination of the QCD spectral densities $\rho^1_{QCD}(s)$ and $\rho^0_{QCD}(s)$,
 $\sqrt{s}\rho_{QCD}^1(s)-m_c\rho_{QCD}^0(s)$, which also leads to a relation between the pole residues $\lambda_{-}\approx 1.5 \lambda_{+}$.   From Table 1 and Fig.1, we can see that in the Borel window in the QCD sum rules for the $M_{+}$, the dominant contributions come from the perturbative term plus the quark condensate term, the higher dimensional vacuum condensates play a minor important role, which is
  in contrast to the QCD sum rules for the $M_{-}$.    So the extracted mass $M_{+}$ is less sensitive to the energy scale $\mu$ of the QCD spectral density than the extracted mass $M_{-}$, see Fig.4.

\begin{figure}
 \centering
 \includegraphics[totalheight=5cm,width=7cm]{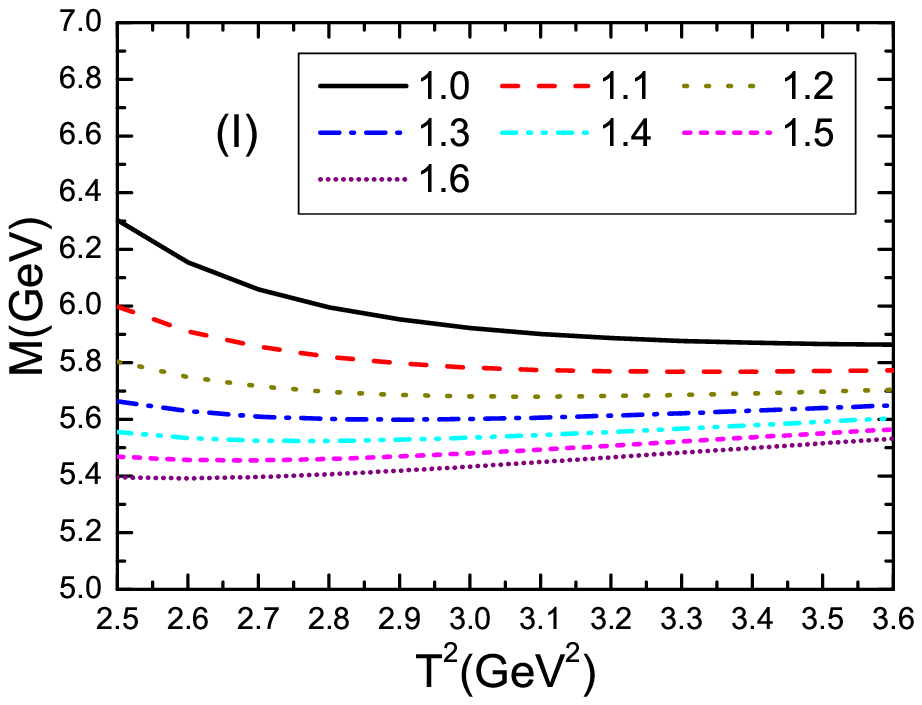}
 \includegraphics[totalheight=5cm,width=7cm]{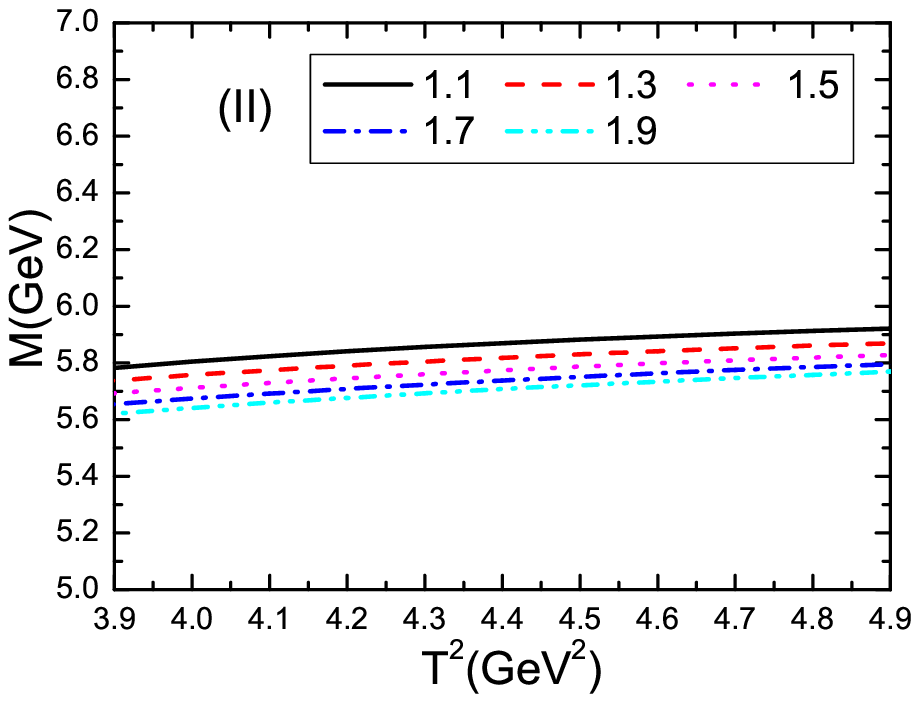}
          \caption{ The masses of the triply-charmed pentaquark states with variations of the Borel parameters $T^2$ and energy scales $\mu$, where  the (I) and (II) denote the negative parity and positive parity  pentaquark states, respectively, the $1.0$, $1.1$, $1.2$, $\cdots$ denote the energy scales $\mu$ of the QCD spectral densities.  }
\end{figure}

The  triply-charmed pentaquark states $P_{cc\bar{c}ud}$ can be produced in the $pp$ collisions at the Large Hadron Collider,
\begin{eqnarray}
pp&\to&\Xi_{bcu}^+\,X\to P_{cc\bar{c}ud}^+ K^- \pi^+\,X\, , \nonumber\\
pp&\to&\Xi_{bbu}^0\,X\to P_{cc\bar{c}ud}^+ K^- \pi^+\pi^-\,X\, ,
\end{eqnarray}
through the decays $b \to c \bar{c}s$ and $b \to c \bar{u} d$ at the quark level, where the superscript $+$ of the $P_{cc\bar{c}ud}^+$ denotes the electronic charge.  The  triply-charmed pentaquark states $P_{cc\bar{c}ud}$ can also be produced in the $\Lambda_{Q}\Lambda_{Q^\prime}$ fusions \cite{Rosner-Nature},
\begin{eqnarray}
\Lambda_b \Lambda_c&\to&\Xi_{bcu}^+\,n\to P_{cc\bar{c}ud}^+ K^- \pi^+\,n\, , \nonumber\\
\Lambda_b \Lambda_b&\to&\Xi_{bbu}^0\,n\to P_{cc\bar{c}ud}^+ K^- \pi^+\pi^- \,n\, .
\end{eqnarray}
We can search for the $P_{cc\bar{c}ud}$ states in their two-body strong decays in the future.

In the following, we perform Fierz re-arrangement  to the current $J(x)$  both in the color and Dirac-spinor  spaces to   obtain the result,
\begin{eqnarray}
J &=&-\frac{1}{4}\mathcal{S}\gamma_5 c\,\bar{c} d+\frac{1}{4}\mathcal{S}\gamma_\lambda \gamma_5 c\,\bar{c}\gamma^\lambda d+\frac{1}{8}\mathcal{S}\sigma_{\lambda\tau} \gamma_5 c\,\bar{c}\sigma^{\lambda\tau} d+\frac{1}{4}\mathcal{S}\gamma_\lambda  c\,\bar{c}\gamma^\lambda\gamma_5 d+\frac{1}{4}i\mathcal{S} c\,\bar{c}i\gamma_5 d\nonumber\\
&&+\frac{1}{4}\mathcal{S}\gamma_5 d\,\bar{c} c-\frac{1}{4}\mathcal{S}\gamma_\lambda \gamma_5 d\,\bar{c}\gamma^\lambda c-\frac{1}{8}\mathcal{S}\sigma_{\lambda\tau} \gamma_5 d\,\bar{c}\sigma^{\lambda\tau} c-\frac{1}{4}\mathcal{S}\gamma_\lambda  d\,\bar{c}\gamma^\lambda\gamma_5 c-\frac{1}{4}i\mathcal{S} d\,\bar{c}i\gamma_5 c \, ,
\end{eqnarray}
 and introduce   the notations  $\mathcal{S}\Gamma c=\varepsilon^{ijk}u^T_i C \gamma_5 c_j \Gamma c_k$ and $\mathcal{S}\Gamma d=\varepsilon^{ijk}u^T_i C \gamma_5 c_j \Gamma d_k$ for simplicity, here the $\Gamma$ denotes the Dirac matrixes.

 The components $\mathcal{S}(x)\Gamma c(x) \bar{c}(x)\Gamma^{\prime}d(x)$ and $\mathcal{S}(x)\Gamma d(x) \bar{c}(x)\Gamma^{\prime}c(x)$ couple potentially to the baryon-meson pairs. The revelent thresholds are
 $M_{\eta_c \Lambda_c^+}=5.270\,\rm{GeV}$,
 $M_{\eta_c \Lambda_c^+(2595)}=5.576\,\rm{GeV}$,
  $M_{\eta_c \Sigma_c^+(2455)}=5.436\,\rm{GeV}$,
   $M_{\eta_c \Sigma_c^+(2520)}=5.501\,\rm{GeV}$,
 $M_{J/\psi \Lambda_c^+}=5.383\,\rm{GeV}$,
  $M_{J/\psi \Lambda_c^+(2595)}=5.689\,\rm{GeV}$,
 $M_{J/\psi \Sigma_c^+(2455)}=5.550\,\rm{GeV}$,
 $M_{J/\psi \Sigma_c^+(2520)}=5.614\,\rm{GeV}$  \cite{PDG},
 $M_{\Xi_{cc}^{++} D^{-}}=5.491\,\rm{GeV}$,
 $M_{\Xi_{cc}^{++} D^{*-}}=5.632\,\rm{GeV}$ \cite{LHCb-Xicc}. After taking into account the currents-hadrons duality, we obtain the Okubo-Zweig-Iizuka super-allowed decays,
\begin{eqnarray}
P\left( {\frac{1}{2}}^-\right) &\to& \eta_c \Lambda_c^+  \, , \, \eta_c \Lambda_c^+(2595)\, , \, \eta_c \Sigma_c^+(2455) \, , \,\eta_c \Sigma_c^+(2520)\, , \, J/\psi \Lambda_c^+\, ,\, J/\psi \Sigma_c^+(2455) \, , \nonumber\\
&&\,\Xi_{cc}^{++} D^{-}\, , \\
P\left( {\frac{1}{2}}^+\right) &\to& \eta_c \Lambda_c^+  \, , \, \eta_c \Lambda_c^+(2595)\, , \, \eta_c \Sigma_c^+(2455) \, , \,\eta_c \Sigma_c^+(2520)\, , \, J/\psi \Lambda_c^+\, ,\,J/\psi \Sigma_c^+(2455) \, ,\,\nonumber \\
&&  J/\psi \Lambda_c^+(2595)\, , \, J/\psi \Sigma_c^+(2520)\, ,\, \Xi_{cc}^{++} D^{-}\, ,\, \Xi_{cc}^{++} D^{*-}\, ,
\end{eqnarray}
we can search for the triply-charmed pentaquark states $P_{cc\bar{c}ud}$ in those decays  in the future.
The LHCb collaboration observed the $P_c(4380)$ and $P_c(4450)$ in the process,
\begin{eqnarray}
pp&\to&\Lambda_b X\to P_c(4380/4450)^+K^-\to J/\psi p K^- X\, ,
\end{eqnarray}
in the $J/\psi p$ invariant mass-spectrum \cite{LHCb-4380}.
The triply-charmed pentaquark states can be observed analogously, for example, in the process
\begin{eqnarray}
pp&\to&\Xi_{bcu}^+\,X\to P_{cc\bar{c}ud}^+ K^- \pi^+\,X\to J/\psi \Lambda_c^+ K^- \pi^+\,X\, ,
\end{eqnarray}
in the $J/\psi \Lambda_c^+$ invariant mass-spectrum. The LHCb collaboration have observed the doubly-charmed baryon state $\Xi_{cc}^{++}$ in the
$\Lambda_c^+K^-\pi^+\pi^+$  invariant mass-spectrum \cite{LHCb-Xicc}, the triply-charmed pentaquark states $P_{cc\bar{c}ud}$ may be observed in the future. 

\section{Conclusion}
In this article, we construct the scalar-diquark-scalar-diquark-antiquark type current to interpolate the  ground state triply-charmed  pentaquark states with $J^P={\frac{1}{2}}^\pm$, and   carry out the operator product expansion up to the vacuum condensates of dimension 10 consistently.
We obtain the QCD spectral densities through dispersion relation and separate the contributions of the negative-parity and positive parity triply-charmed  pentaquark states explicitly. Then we extract the masses and pole residues in the Borel windows at the optimal energy scales of the QCD spectral densities, which are determined by  the energy scale formula $\mu=\sqrt{M^2_{P}-(3{\mathbb{M}}_c)^2}$.
Experimentally, there are no candidates for the triply-charmed pentaquark states,  we can search for the triply-charmed  pentaquark states in the  Okubo-Zweig-Iizuka  super-allowed strong decays  in the future.

\section*{Acknowledgements}
This  work is supported by National Natural Science Foundation, Grant Number  11775079.

\end{document}